\documentstyle[12pt,aaspp4]{article}

\def\lax    {\ifmmode{_<\atop^{\sim}}\else{${_<\atop^{\sim}}$}\fi}
\def\gax    {\ifmmode{_>\atop^{\sim}}\else{${_>\atop^{\sim}}$}\fi}

\def\farcs{\hbox{$.\!\!^{\prime\prime}$}} 
 
\def\pagebreak{\vfill\eject}

\def\Msun{{M$_\odot$}}
\def\Lsun{{L$_\odot$}}

\def\sun{\mbox{$_\odot$}}

\def\deg{{$^\circ$}}
\def\half{{\leavevmode\kern.1em\raise.5ex
\hbox{\the\scriptfont0 1}\kern-.1em /
\kern-.15em\lower.25ex\hbox{\the\scriptfont02}}} 
\def\gtsim{\lower.5ex\hbox{$\buildrel > \over\sim$}}
\def\ltsim{\lower.5ex\hbox{$\buildrel < \over\sim$}}
\def\sun{\mbox{$_\odot$}}
\def\kms{~km~s$^{-1}$}

\def\cof{CO(J=1$-$0)}
\def\tco{$^{13}$CO}
\def\tcof{$^{13}$CO(J=1$-$0)}

\def\h{H$_2$}

\def\water{H$_2$O}

\def\jyb{~Jy~beam$^{-1}$}
\def\jybkms{~Jy~beam$^{-1}$ km~s$^{-1}$}

\def\mjyb{~mJy~beam$^{-1}$}
\def\ra#1#2#3{#1$^{\rm h}$#2$^{\rm m}$#3$^{\rm s}$}
\def\dec#1#2#3{$#1^\circ#2'#3''$}

\slugcomment{ApJ, (to be published in v584 n2 February 20, 2003)}

\begin{document}

\title{
Clustered Star Formation in W75~N}

\author{
D.S.~Shepherd\altaffilmark{1}, 
L. Testi\altaffilmark{2}, \&
D. P. Stark\altaffilmark{3}
}

\vspace{-3mm}
\altaffiltext{1}{National Radio Astronomy Observatory, P.O. Box 0,
Socorro, NM 87801\\
The National
Radio Astronomy Observatory is a facility of the National Science
Foundation operated under cooperative agreement by Associated
Universities, Inc.}
\altaffiltext{2}{Osservatorio Astrofisco di Arcetri, Largo Enrico
Fermi 5, I-50125 Firenze}
\altaffiltext{3}{Department of Astronomy, University of Wisconsin --
Madison, 475 N. Charter St., Madison, WI 53706}

\rightskip=\leftskip 
\vspace{-3mm}

\begin{abstract}

We present $2''$ to $7''$ resolution 3~mm continuum and \cof\ line
emission and near infrared K$_s$, \h, and [FeII] images toward the
massive star forming region W75~N.  The CO emission uncovers a complex
morphology of multiple, overlapping outflows.  A total flow mass of $>
255$~M\sun\ extends 3~pc from end-to-end and is being driven by at
least four late to early-B protostars.  More than 10\% of the
molecular cloud has been accelerated to high velocities by the
molecular flows ($> 5.2$\kms\ relative to $v_{LSR}$) and the
mechanical energy in the outflowing gas is roughly half the
gravitational binding energy of the cloud.  The W75~N cluster members
represent a range of evolutionary stages, from stars with no apparent
circumstellar material to deeply embedded protostars that are actively
powering massive outflows. Nine cores of millimeter-wavelength
emission highlight the locations of embedded protostars in W75~N.  The
total mass of gas \& dust associated with the millimeter cores ranges
from 340~M\sun\ to 11~M\sun.  The infrared reflection nebula and
shocked \h\ emission have multiple peaks and extensions which, again,
suggests the presence of several outflows.  Diffuse \h\ emission
extends about 0.6 parsecs beyond the outer boundaries of the CO
emission while the [FeII] emission is only detected close to the
protostars.
The infrared line
emission morphology suggests that only slow, non-dissociative J-type
shocks exist throughout the pc-scale outflows.  Fast, dissociative
shocks, common in jet-driven low-mass outflows, are absent in W75~N.
Thus, the energetics of the outflows from the late to early B
protostars in W75~N differ from their low-mass counterparts -- they do
not appear to be simply scaled-up versions of low-mass outflows.

\end{abstract}

\vspace{-.5cm}

\section{INTRODUCTION}

W75~N is a massive star forming region with an integrated IRAS
luminosity of $1.4 \times 10^5$~\Lsun\ (Moore, Mountain, \& Yamashita
1991; Moore et al. 1988,1991).  The W75~N cloud is located at a
distance of 2~kpc (Dickel, Wendker, \& Bieritz 1969), just $15'$ north
of the massive outflow system DR~21 powered by a cluster of OB stars
(e.g. Garden et al. 1991 and references therein).  Both DR~21 and
W75~N are part of the Cygnus-X complex of dense molecular clouds.
Haschick et al. (1981) identified three regions of ionized gas in
W75~N at a resolution of $\sim 1.5''$: W75~N (A), W75~N (B), and W75~N
(C).  Hunter et al. (1994) later resolved W75~N (B) with $\sim 0.5''$
resolution into three regions: Ba, Bb, and Bc.  Torelles et al. (1997)
then imaged W75~N (B) at $\sim 0.1''$ resolution, and detected Ba and
Bb (which they called VLA~1 \& VLA~3), along with another weaker, and
more compact HII region, VLA~2.

A parsec-scale molecular outflow originates near the cluster of
ultracompact HII (UC HII) regions in W75~N (B).  The mass of the CO
outflow has been estimated to be 50~M\sun\ to 500~M\sun\ based on
single-dish, CO observations (e.g Fischer et al. 1985; Hunter et
al. 1994; Davis et al. 1998a,b; Ridge \& Moore 2001).  The UC HII
regions have a combined $L_{bol}$ of $4.4 \times 10^4$~L\sun\ and most
are in a protostellar phase based on the presence of OH, \water, \&
methanol masers, and compact millimeter continuum emission (Baart et
al. 1986; Hunter et al. 1994; Torrelles et al. 1997; Minier, Conway,
\& Booth 2000, 2001; Shepherd 2001; Hutawarakorn, Cohen, \& Brebner
2002; Slysh et al. 2002; Watson et al. 2002).  Several studies have
assumed the flow is dominated by a single massive star: the central
source in the UC HII region VLA~1 (Ba) because the position angles of
the ionized gas and the CO emission are similar (Hunter et al. 1994;
Torrelles et al. 1997; Davis et al. 1998a,b).  Shepherd (2001)
suggested VLA~3 (Bb) and, perhaps, VLA~2 may be the primary powering
sources based on the presence of compact millimeter continuum
emission.  More recently, Hutawarakorn et al. (2002) suggested VLA~2
is the dominant source powering the outflow based on OH maser
emission.  Given the sheer number of interpretations, it is clear that
W75~N is a confused region.

Assuming a primary driving source for the CO outflow, Davis et
al. (1998b) suggested that the CO red-shifted lobe and \h\ morphology
supported a jet-driven, bow-shock entrainment scenario in which a
steady, over-dense molecular jet, developed to explain
highly-collimated outflows from low-mass protostars, was applied to
W75~N (Lada \& Fich 1996; Smith et al. 1997; \& Suttner et al. 1997).
The proposed model implied a jet radius of 0.03~pc at 1.3~pc from the
star with a jet opening angle of about 2.6\deg\ (Richer et al. 2000).
If a powerful, well-collimated jet was being driven by an OB protostar
in W75~N, it would provide strong constraints on outflow/accretion
theories for luminous protostars (see, e.g., Shang et al. 2002;
Cabrit, Ferreira, \& Raga 1999; K\"onigl 1999; Shu et al. 2000;
K\"onigl \& Pudritz 2000).

To obtain a better understanding of the number of sources driving
outflows and the energetics of the flow(s), we have made
interferometric mosaics of the W75~N region in CO(J=1--0) and
millimeter continuum using the Owens Valley Radio Observatory and
obtained images at near-infrared wavelengths using the Telescopio
Nazionale Galileo to compare the morphology of the shocked gas \&
infrared nebulosity with the CO emission.

\section{OBSERVATIONS}

\subsection{Owens Valley Observations in the 3~mm band}

Observations in 2.7~mm continuum and {\cof} line were made with the
Owens Valley Radio Observatory (OVRO) array of six 10.4~m telescopes
between 1999 March 15 and 1999 December 4.  Projected baselines
ranging from 15 to 115 meters provided sensitivity to structures up to
about $16''$.  The final $\sim 5' \times 1.5'$ mosaic images of both
line and continuum emission are made up of 17 fields with primary
beam $65''$ (FWHM) spaced $30''$ apart.  The total integration time on
source was approximately 3.25 hours/pointing center.  Cryogenically
cooled SIS receivers operating at 4~K produced typical single sideband
system temperatures of 200 to 600~K.  The gain calibrator was the
quasar BL Lac and the bandpass calibrators were 3C~454.3 and 3C~345.
Observations of Uranus, Neptune, or 3C~273 provided the flux density
calibration scale with an estimated uncertainty of $\sim 20$\%.
Calibration was carried out using the Caltech MMA data reduction
package (Scoville et al. 1993).  Images were produced using the MIRIAD
software package (Sault et al. 1995) and deconvolved with a
maximum-entropy-based algorithm designed for mosaic images (Cornwell
\& Braun 1988).

The CO $uv$ data at 115.27~GHz were convolved with a $5''$ taper
resulting in a synthesized beam of $6.46'' \times 6.28''$ (FWHM) at
P.A. $-54.7$\deg.  The spectral resolution was 2.6{\kms} and the RMS
noise was 0.13\jyb.  The spectral band pass was centered on the local
standard of rest velocity ($v_{LSR}$) of $10.0$~\kms\ (the assumed
systemic velocity of the W75~N cloud), taken from the CS(J=7--6)
emission peak (Hunter et al. 1994).  Simultaneous 2.7~mm continuum
observations were made in a 1~GHz bandwidth channel with central
frequency 112.77~GHz.  The $uv$ data were convolved with a $6''$ taper
resulting in a synthesized beam $7.29'' \times 7.13''$ (FWHM) at
P.A. $-62.9^\circ$.  The RMS noise was 3.6~mJy~beam$^{-1}$.

An additional on-source integration time of 4.6 hours was obtained
with OVRO centered on the position of W75~N:MM~1 ($\alpha(J2000) =$
\ra{20}{38}{36.36} $\delta(J2000) = $ \dec{42}{37}{33.5}).
Observations were made on 1999 March 29 and 2001 March 18.  Baselines
between 35 and 240 meters provided sensitivity to structures up to
$6''$.  The CO $uv$ data (spectral resolution 2.6\kms) were convolved
with a $1.5''$ taper resulting in a synthesized beam of $2.04'' \times
1.77''$ (FWHM) at P.A. $-80.5$\deg.  The final RMS noise was
50~mJy~beam$^{-1}$ in each channel.

\subsection{3~mm single dish spectra}

Observations were made in \cof\ and \tcof\ with the Kitt Peak 12~m
telescope on 2000 May 9 using the SIS 3~mm receiver with 1~MHz filter
banks centered on $v_{LSR} = 10.0$~\kms\ to give a velocity resolution
of 2.6\kms\ and a total bandwidth of 650\kms.  System temperatures
ranged from 230~K for \tco\ to 360~K for \cof.  The
half-power beam width (HPBW) at 115 GHz is about $60''$.  

Single dish spectra were obtained at three positions (J2000
coordinates): centered on W75~N:MM~1 ($\alpha =$~\ra{20}{38}{36.50}
$\delta =$~~ \dec{42}{37}{33.5}); and in the south-east and north-west
outflow lobes ($\alpha =$~\ra{20}{38}{31.00}~~ $\delta
=$~\dec{42}{36}{51.0} and $\alpha =$~\ra{20}{38}{41.00}~~ $\delta
=$~\dec{42}{37}{51.0}, respectively).  The data were reduced with the
NRAO UniPOPS software package.  The resulting spectra were used to
estimate the optical depth in \cof\ as a function of velocity and
position in the W75~N region.

\subsection{Near-Infrared Observations}

Near-infrared observations of the W75~N region were made on 2000 June
16, at the 3.5~m Telescopio Nazionale Galileo (TNG) at the Roque de
Los Muchachos Observatory on the Spanish island of La Palma.  The
ARNICA NIR imager (Lisi et al.~\cite{Lea96}; Hunt et al.~\cite{Hea96})
was used to obtain images in the H$_2$ narrow band filter and in the
K$_s$ broad band filter. ARNICA is equipped with a HgCdTe $256 \times 256$
NICMOS3 infrared array.  The pixel scale, when coupled with the TNG,
is 0.355~arcsec/pixel, and the corresponding field of view is
$1.5 \times 1.5$~arcmin$^2$ per frame. The seeing at the time of the
observations was about 0\farcs 9.

To search for \h\ emission beyond the CO emission, a 14~pointing
mosaic pattern was employed covering an area of approximately
9$\times$2~arcmin$^2$ that was roughly aligned with the CO flow. The
mosaic was repeated several times, dithering the telescope by a few
pixels each time, until the desired integration time was achieved. The
final integration times per sky position was 8 minutes in K$_s$-band
and 30~minutes in the H$_2$ filter.

Data reduction and analysis were performed using the IRAF software
package.  Following standard flat-fielding and sky subtraction, the
individual images were registered and the final mosaic was
produced. The K$_s$-band observations were calibrated using standard
stars from the ARNICA list (Hunt et al.~\cite{Hea98}). The H$_2$
mosaic was calibrated assuming that a set of stars have the same flux
density in the narrow- and broad-band filters.  The broad-band mosaic
was then used to subtract the continuum emission from the H$_2$
mosaic. Integrated line fluxes are then estimated assuming the width
of the H$_2$ filter as measured by Vanzi et
al.~(\cite{Vea98}).\footnote{Due to the different optical
configuration used at the TNG with respect to that used at the TIRGO
telescope, the narrow-band filters do not suffer from the effective
field of view reduction discussed by Vanzi et al.~(\cite{Vea98}).}
The calibration accuracy is expected to be within 20\%.  Accurate
($\le 0\farcs 5$) astrometry was derived for both mosaics using
stellar positions from the 2MASS second incremental data release.

Additional observations of two $\sim$4$^{\prime}\times4^{\prime}$
fields centered north-east and south-west of W75~N were obtained in
2002 August 21 using the TNG near infrared camera spectrograph (NICS,
Baffa et al.~\cite{Bea01}). Each of the two fields was observed
through the H$_2$~($\lambda$=2.12~$\mu$m) and
[FeII]~($\lambda$=1.64~$\mu$m) narrow-band filters and in two
narrow-band continuum filters, K$_{cont}$ and H$_{cont}$; a detailed
characterization of all these filters can be found in Ghinassi et
al.~(\cite{Gea02}). The observations were reduced and astrometrically
calibrated following the procedure outlined above.  The weather
conditions were not photometric during the observations so the data
could not be flux calibrated.  Line-only images were obtained by
subtracting the narrow-band continuum images from the line$+$continuum
images.  The subtraction was not perfect on strong stellar sources or
on stars with very red or very blue spectra.

\section{RESULTS}

\subsection{\h\ and [FeII] morphology}

Infrared reflection nebulosity is associated with two distinct regions
of ionized gas (Fig. 1): W75~N~(A) at position $\alpha(J2000) =$
\ra{20}{38}{38}~ $\delta(J2000) =$ \dec{42}{37}{59} and W75~N~(B) at
position $\alpha(J2000) =$ \ra{20}{38}{37}~ $\delta(J2000) =$
\dec{42}{37}{32} (e.g. Haschick et al. 1981, Moore et al. 1988).
Shock-excited \h\ emission (Figs. 2 \& 3) is present in the north-east
(near $\alpha(J2000) =$ \ra{20}{38}{52}~~ $\delta(J2000) =$
\dec{42}{39}{00}) and along the CO emission boundaries (see also Davis
et al. 1998a,b).  Our \h\ images also show that faint, patchy \h\
emission extends nearly an arcminute (0.6~pc at a distance of 2~kpc)
beyond the south-west CO flow ($\alpha(J2000) =$ \ra{20}{38}{22--18}~~
$\delta(J2000) =$ \dec{42}{36}{10}).

The continuum subtracted H$_2$ mosaic is shown in Fig.~2.  Following
the nomenclature used by Davis et al.~(1998a,b) for the south-west
portion of the H$_2$ flow, all previously known H$_2$ knots and
filaments are marked with solid lines.  These are labeled SW--A to
SW--H in the south-west flow, C--A and C--B in the central region,
and NE--A to NE--F in the north-east flow.  Additionally, Figs. 2 \& 3 
reveal faint diffuse H$_2$ emission beyond the tip of the
south-west flow and within the north-east flow. These new features are
marked with dashed lines and labeled SW--I to SW--K and NE--G to
NE--I.  Faint, diffuse H$_2$ features as well as the filamentary
structure of the emission in knots NE--D, NE--E, \& NE--F are confirmed
by observations obtained in 2002 August.  Figure 3 presents an overlay
of the continuum subtracted [FeII] emission (contours) on the H$_2$
(greyscale) images from the 2002 August observations.  [FeII] line
emission is only detected close to the UC HII regions in W75~N B and
near the exciting star of W75~N A.  No [FeII] emission is detected in
the outer flow regions.  The non-detection of [FeII] far from the
protostars does not appear to be due to higher extinction since we
clearly detect 2.12$\mu$m \h\ emission beyond the CO outflow boundaries
and we detect [FeII] emission near the cloud core where the column
density is higher.

A chain of H$_2$ knots, apparently unrelated to the main flow, are
detected near $\alpha(J2000) =$ \ra{20}{38}{32 -- 36}~~ $\delta(J2000)
=$ \dec{42}{39}{30-60} (Fig. 3).  This jet is outside of the CO and
infrared mosaic fields.  The knots may be associated with a jet from a
young star north of the main complex discussed in this paper.

\subsection{CO outflows and their driving sources}

A high-velocity CO outflow, centered near W75~N (B), measures 3~pc
from end-to-end (projected length) and extends well beyond the
infrared reflection nebula (Figs. 1 \& 4).  Red and blue-shifted CO
emission exists both in the north-east and south-west.  The CO mosaic
did not include areas to the north-west and south-east so it is
unclear if high-velocity CO exists in these regions.  The boundaries
and flux density of the CO outflow are well determined on the
red-shifted side of the line, however, at velocities between 0 and
--5.6~\kms\ the DR~21 cloud ($v_{LSR} = -2.5$~\kms) confuses the
identification of the outflow structure.

Nine millimeter continuum peaks showing the locations of warm dust
emission are identified in Fig. 5 \& Table 1 (W75~N:MM 1 through
W75~N:MM 9\footnote{Names of millimeter cores are shortened to MM 1 to
MM 9 for the remainder of this paper.}).  MM~1 -- MM~4 are near the
origin of the outflow activity and lie $5-10''$ from the W75~N~(B)
reflection nebulosity to the north and west.  MM~5 is associated with
the more extended HII region, W75~N~(A), while MM~6 -- MM~9 are not
associated with any previously known sources.  Infrared counterparts
do not exist for the millimeter sources (except MM~5) suggesting that
these sources are too deeply embedded to be detected at 2$\mu$m.
Figure 5 also shows the $2''$ resolution image from Shepherd (2001)
for comparison.  The $2''$ resolution resolved the individual
millimeter cores MM~1--4 but resolved out the more extended emission
associated with MM~5 and MM~6.  Millimeter cores MM~7--9 were outside
of the primary beam of the Shepherd (2001) observations and thus, were
not detected.  Figure 5 also compares the MM~5 millimeter source with
narrow-band \h$+$continuum emission and K$_s$ broad-band emission in
W75~N~(A).  The central star of W75~N~(A) (spectral type B0.5;
Haschick et al. 1981) is clearly visible in the infrared and is
surrounded by a $20''$ shell of thermal dust emission.  Diffuse
reflection nebulosity is centered on the star and a wisp of \h\
emission is visible just east of the star ($\alpha(J2000) =$
\ra{20}{38}{38--39}~~ $\delta(J2000) =$ \dec{42}{38}{00}).  

Compact, high-velocity CO emission appears to originate from the
UC~HII regions VLA~1 (Ba) \& VLA~3 (Bb) and from MM~2 (Figs. 6, 7, \&
8).  The $2 - 5''$ resolution is not sufficient to determine if VLA~2,
located only $0.5''$ north of VLA~3 (Bb), is also associated with
high-velocity CO gas.  A detailed discussion of each of the proposed
outflows is given below.

{\bf The outflow from VLA~1 (Ba):} VLA~1 (Ba) is a thermal jet source
associated with \water\ and OH masers (Baart et al. 1986, Torrelles et
al. 1997).  It is embedded in the MM~1 core detected in 1 \& 3~mm
continuum emission (Shepherd 2001).  The spectral type of the powering
source is unknown since the observed centimeter continuum emission is
likely due to the ionized jet rather than emission from an
ionization-bounded UC HII region. Red-shifted emission to the
north-east of MM~1 can be traced to the jet-like ionized flow from
VLA~1~(Ba).  Figure 7 presents a $6''$ image of the integrated CO
emission from the flow as well as a position-velocity diagram from a
slice along the proposed flow axis (P.A. 51\deg).  A ridge of CO
emission extends to the north-east with projected velocity greater
than 25\kms\ (relative to $v_{LSR}$) almost an arcminute from the UC
HII region.  High-velocity CO emission is also centered on the UC HII
regions (position offset $-18''$ in the PV diagram of Fig. 7).  A
$2''$ resolution image (Fig. 6, top left) shows that this red-shifted
emission near the base of the outflow appears to be produced by VLA~1
(Ba) as well as VLA~3 (Bb) and possibly VLA~2.  The $2''$ resolution
of Fig. 6 resolves out much of the extended emission in the flow
leaving only compact clumps visible along the flow axis.  The P.A. is
similar to the elongation of the ionized emission in VLA~1 (Ba)
(P.A. $\sim 43$\deg) and a line of {\water} masers detected along the
jet (Torrelles et al. 1997).  The molecular flow is also seen in
Fig. 1 as a well-collimated, red-shifted lobe extending to the
north-east from the MM~1 core.  The outflow is shown as one-sided in
Figs. 6 \& 7 because only one side of the flow is detected.  The
counterflow may exist however it may be too extended to image at this
high-resolution or it may be expanding into a less dense medium that
would not create appreciable CO emission.

{\bf The outflow from VLA~3 (Bb):} VLA~3 (Bb) is a compact UC HII
region in MM~1 with a central star of spectral type B0.5 to B0.  It is
associated with a single \water\ maser and compact 1 \& 3~mm continuum
emission.  A lower limit on the mass of warm gas and dust within 2000~AU of
the protostar is 5~M\sun\ (Shepherd 2001).  Figure 6 illustrates that
compact red-shifted emission to the east and blue-shifted emission to
the west of MM~1 can be traced to VLA~3 (Bb) with P.A. $\sim 101$\deg.
There are no obvious features in the extended emission that correspond
to an outflow with this orientation, however, the CO mosaic did not
extend to the north-west or south-east so a large-scale outflow could
have been missed (Fig. 1).  The molecular gas morphology does not seem
to be correlated with that of the ionized gas or {\water} masers near
the source: the ionized gas is slightly elongated along P.A. 149\deg\
and the \water\ maser is located near the southern boundary of the
ionized gas.

{\bf The outflow from MM~2:} MM~2 is a molecular core
identified by compact, warm dust emission at 1 \& 3~mm and \water\
maser emission (Torrelles et al 1997, Shepherd 2001).  The mass of the
core is $30-50$~M\sun.  No ionized gas has been detected indicating
that the spectral type is less than a B2 star or high accretion is
preventing the formation of a UC HII region.  High-velocity,
blue-shifted emission (13 to 36\kms\ projected velocity relative to
$v_{LSR}$) can be traced from the MM~2 core to the south-east
(P.A. 124\deg).  Figure 8 presents a $6''$ image of the integrated
emission from the flow as well as a position-velocity diagram from a
slice along the proposed flow axis.  The velocity of the jet relative
to $v_{LSR} = 10$\kms\ increases away from the position of MM~2 and
remains well collimated.  Diffuse emission at velocities greater than
$v = -5$\kms\ is due to the W75~N \& DR~21 clouds.  Figure 6  
shows the more compact emission in the flow.  Three clumps
of high-velocity gas are detected extending away from MM~2 along with
faint emission at the location of the core.  The compact clumps appear
to trace a shell of dense gas surrounding the outflow axis.  The
molecular outflow is identified as one-sided because the red-shifted
counterflow was not detected.  

The diffuse millimeter core MM~4 is located along the axis of the
proposed MM~2 outflow.  Although no \water\ or OH maser emission has
been detected toward MM~4, nor has centimeter or 1~mm continuum
emission been detected, there is diffuse warm dust emission traced by
3~mm continuum.  Thus, MM~4 may harbor an embedded protostar.
Assuming the MM~4 core is heated internally, the lack of maser
activity, compact warm dust emission at 1~mm, and the absence of
ionized gas emission at centimeter wavelengths suggest that MM~4 is a
low-mass protostar (Shepherd 2001).  With the current resolution and
sensitivity, we can not determine if the MM~4 protostar is
contributing to the observed flow dynamics.

The proposed position angles of the outflowing gas (illustrated by
arrows in Figs. 6, 7, \& 8) are 51\deg\ for VLA~1 (Ba), 101\deg\ for
VLA~3 (Bb), and 124\deg\ for MM~2.  The position angle of the
parsec-scale outflow detected in \h\ and CO (Fig. 1) is 62.5\deg.
Although the orientation of the VLA~1 (Ba) outflow is similar to that
of the parsec-scale flow, it does not appear likely that VLA~1 (Ba) is
the powering source.  Assuming the VLA~1 (Ba) flow is
symmetric, a blue-shifted counterflow is expected in the south-west,
not a red-shifted flow.  Thus, our observations do not not identify
the source responsible for the 3~pc outflow which dominates the
large-scale morphology and kinematics of the region.

\subsection{Mass \& Kinematics of the Outflows} 

The mass associated with CO line emission is calculated following the
method proposed by Scoville et al. (1986).  The CO excitation
temperature near the millimeter continuum emission varies from about
35 to 75~K with $\sim 50$~K being the median value near the MM~1 peak
(Davis et al. 1998b).  Rotational temperatures derived from CH$_3$CN
in the cloud core vary from 47 to 78~K, consistent with the Davis et
al. estimates (Kalenski\u{\i} et al. 2000).  We assume the gas is in
LTE, at a temperature of 50~K, with [CO]/[\h]~=~$10^{-4}$, and
[CO]/[\tco]~=~71 at the galacto-centric distance of 8.5~kpc (Wilson \&
Rood 1994).  The CO optical depth as a function of velocity and
position is calculated using single dish \cof\ and \tcof\ spectra
taken at three positions within the W75~N region (Fig. 9).  We assume
\tco\ is optically thin at all velocities which is probably valid in
the line wings; however, \tco\ is likely to be optically thick near
the line core.  In channels where no {\tco} emission is detected, we
assume the CO is optically thin.  The CO channel images (Fig. 4) show
that the emission near $v_{LSR}$ is almost entirely resolved out by
the interferometer.  If high-velocity ($> 5.2$\kms) structures exist
that are larger than the largest angular scale that can be imaged ($>
16''$), then our mass estimate represents a lower limit.

Because multiple, overlapping flows are present, it is not possible to
obtain a reasonable estimate of the inclination of each flow.  Thus,
we assume an inclination angle of 45\deg\ which minimizes errors
introduced by inclination effects.  Table 2 summarizes the physical
properties of the molecular gas in the combined outflows originating
within MM~1 and from the blue-shifted MM~2 outflow lobe.  The total
flow mass $M_f$ is given by $\sum M_i$ where $M_i$ is the flow mass in
velocity channel $i$ corrected for optical depth.  The momentum $P$ is
given by $\sum M_i v_i$ and the kinetic energy E by $\frac{1}{2} \sum
M_i v_i^2$ where $v_i$ is the central velocity of the channel relative
to $v_{LSR}$.  The characteristic flow timescale $t_d$ is $R_{f}/<V>$,
where the intensity-weighted velocity $<V>$ is given by $P/(\sum M_i)$
(Cabrit \& Bertout 1990) and $R_{f}$ is the flow radius.  The mass
outflow rate $\dot M_{f}$ is $\sum M_i/t_d$ and the force F is
$P/t_d$.  Assuming D=2~kpc, the total molecular mass in outflowing gas
($v > 5.2$\kms\ relative to $v_{LSR}$) is $> 255$~M{\sun}.  The values
presented in Table 2 are derived assuming all flows have the same
systemic velocity.  This is a reasonable assumption for the driving
sources of the combined MM~1 outflows since the observed UC HII
regions are embedded within the same molecular clump and have a
projected separation of only 0.5 to $1''$ (1000 - 2000 AU at a
distance of 2~kpc).  The source driving the outflow from the MM~2
molecular core has a projected separation of about $5''$ (10,000~AU or
0.05~pc) from MM~1.  It is possible that MM~2 could have a
slightly different systemic velocity from MM~1 that we cannot detect
in our images or single dish spectra.  If the systemic velocity of the
MM~2 core is different by a factor of $\Delta v$ from the assumed
velocity of 10\kms, then the error in the momentum estimates will be
proportional to $|\Delta v|$ and mechanical energy to $|\Delta v^2|$.

The combined MM~1 outflows have a total mass of at least 165~M\sun\ and
energy, E $> 3.4 \times 10^{47}$~ergs.  The MM~2 outflow has a total
mass $> 90$~M\sun, E $> 1.8 \times 10^{47}$~ergs.  The CO mosaic did
not extend in the south-east direction of the MM~2 flow, thus, the
full outflow was not imaged and age and $\dot M$ should be considered
a lower limit.  Despite the uncertainties, the flow masses and
energies are consistent with those for outflows driven by young, early
B stars. This is in agreement with the estimated spectral types of the
stars powering the UC~HII regions in MM~1 (B2 to O9; Hunter et
al. 1994; Torrelles et al. 1997; Shepherd 2001; Slysh et al. 2002).

\subsection{Circumstellar material near embedded protostars}

The mass of gas and dust associated with warm dust being heated by the
central protostars is estimated from the millimeter continuum emission
using ${\rm M}_{gas + dust} = \frac{{\rm F}_{\nu}~ {\rm D}^2} {{\rm
B}_{\nu}({\rm T}_d)~ \kappa_{\nu}}$ where D is the distance to the
source, ${\rm F}_{\nu}$ is the continuum flux density due to thermal
dust emission at frequency $\nu$, and ${\rm B}_{\nu}$ is the Planck
function at temperature T$_d$ (Hildebrand 1983).  Assuming a
gas-to-dust ratio of 100, the dust opacity per gram of gas is taken to
be $\kappa_{\nu} = 0.006(\frac{\nu}{245 {\rm
GHz}})^{\beta}$~cm$^2$~g$^{-1}$ where $\beta$ is the opacity index
(see Kramer et al. 1998; and the discussion in Shepherd \& Watson
2002).  This value of $\kappa$ agrees with those derived by
Hildebrand (1983) and Kramer et al. (1998) to within a factor of 2.
The opacity index $\beta = 1.5$ appears to be appropriate between
wavelengths of 650~microns and 2.7~mm for sub-micron to
millimeter-sized grains expected in warm molecular clouds and young
disks (Pollack et al. 1994).  We assume the emission is optically
thin and the temperature of the dust can be characterized by a single
value.  Using values of $T_d = 50$~K and $\beta = 1.5$, we find the
total mass of gas and dust associated with the 2.7~mm continuum
emission is approximately 475~M{\sun} (Table 1).  Our results are
consistent with those of Shepherd (2001) and Watson et al. (2002) to
within the errors.

\section{DISCUSSION}

The total molecular mass of outflowing gas from the MM~1 and MM~2
combined flows ($v > 5.2$\kms\ relative to $v_{LSR}$) is $>
255$~M{\sun}. Hunter et al. (1994) found $M_f = 48$~\Msun\ with a
rough scaling performed to take into account an optical depth
correction.  However, their image covered only the inner region of the
flow so their estimate should be considered a lower limit.  Based on
single dish observations of CO(J=3--2), Davis et al (1998a) estimated
a total flow mass of $M_f = 272$~\Msun, uncorrected for optical depth
effects.  This mass estimate is extremely high for an optically thin
approximation.  Examination of their Fig. 11, T$_A^*$ vs. ($v-v_o$),
shows that the blue-shifted lobe has an order of magnitude increase in
the integrated flux at the velocity of DR~21 (--2.5\kms).  It appears
that their single-dish map may have been significantly contaminated by
emission from the DR~21 cloud, which introduced uncertainties in the
mass and kinematics estimates.  Ridge \& Moore (2001) estimated the
outflow mass of the red-shifted lobe only to be 273~\Msun\ based on a
CO(J=2--1) single dish image corrected for optical depth.  The mass of
blue-shifted gas was not estimated by Ridge \& Moore due to the
contamination by the DR~21 cloud.  This value is significantly higher
than our estimate and may be due to missing extended emission in the
interferometer image, especially at low velocities.  Despite this
problem, interferometric imaging also provided benefits: it was easier
to distinguish between outflow gas and the DR~21 cloud and to identify
flows from multiple sources in the cluster.

The total cloud core mass of W75~N has been estimated to be
1800--2500~\Msun\ based on observations at submillimeter wavelengths
(Moore, Mountain, \& Yamashita, 1991) and 1200~\Msun\ based on
CS(J=7--6) emission (Hunter et al. 1994).
The gravitational binding energy of the cloud, $G M_{cloud}^2/c_1 r$,
is 1--2$\times 10^{48}$~ergs, where we take the radius
$r=0.25$~pc (Moore et al. 1991) and $c_1$ is a constant which depends
on the mass distribution ($c_1 = 1$ for $\rho \propto r^{-2}$).  
More than 10\% of the molecular cloud is participating in the outflow
and the combined outflow energy is roughly half the gravitation
binding energy of the cloud.  The observed W75~N outflows are
injecting a significant amount of mechanical energy into the cloud
core and may help prevent further collapse of the cloud.

Our CO(J=1--0) images suggest that high-velocity gas is
associated with at least two UC HII regions: VLA~1~(Ba) and
VLA~3~(Bb) and an embedded source in the millimeter core MM~2.  The
position angles of the individuals outflows are not aligned, ranging
from 51\deg\ to 124\deg.  The \h\ morphology is diffuse and patchy
both in the north-east and south-west.  The irregular morphology of
the infrared reflection nebula with fingers of nebulosity radiating
out from the MM~1/MM~2 millimeter cores supports the conclusion that
multiple energetic outflows are carving large cavities in the
molecular cloud.

Low surface brightness \h\ emission extends well beyond the CO outflow
while [FeII] emission is only detected close to the protostellar
cluster.  It is generally believed that [FeII] line emission
associated with low-mass outflows requires the presence of fast,
dissociative shocks that disrupt dust grains and release heavy
elements just behind a Jump-shock (J-shock) boundary.  \h\ emission,
on the other hand, appears to be produced in slow, non-dissociative
J-type shocks (e.g.  Hollenbach \& Mckee 1989; Smith 1994; Gredel
1994; Beck-Winchatz et al. 1996).  Continuous-shocks (C-shocks) cannot
easily produce emission from ionized species such as [FeII] nor can
they produce the observed column densities typically seen in \h\
toward Herbig Haro objects from low-mass protostars (Gredel 1994).  In
fact, Nisini et al. (2002) find that there appears to be no
correlation between \h\ and [FeII] emission in outflows from low-mass
YSOs which supports the interpretation that physically different
mechanisms are responsible for producing \h\ and [FeII] emission.  In
a sample of Herbig Haro objects produced by jets from low-mass
protostars, both \h\ and [FeII] is found toward all sources and the
morphology of \h\ and [FeII] emission is similar on large scales
although it differs in the detail (Gredel 1994; Reipurth et al. 2000).
These observations indicate that jets from low-mass protostars produce
both fast, dissociative regions where [FeII] emissions arises and
slower, non-dissociative regions where \h\ emission arises.  In
contrast, [FeII] emission toward W75~N is only detected close to the
central sources and does not show a jet-like morphology as in outflows
from low- and intermediate-mass young stellar objects (e.g. Lorenzetti
et al. 2002; Nisini et al. 2002; Reipurth et al. 2000). The outflows
from W75~N appear to exhibit only slow, non-dissociative J-type shocks
which produce copious \h\ emission throughout the outflow region but
the fast, dissociative shocks responsible for [FeII] emission are
absent in the outer regions of the flow.  Instead, the diffuse [FeII]
line emission in W75~N is coincident with the brightest K$_s$-band
reflection nebulosity.  One possibility may be that the [FeII]
emission traces photo-dissociation regions (PDRs) along cloud surfaces
illuminated by the massive protostars in the MM~1 core.  This
situation is also observed in the Orion Bar PDR (e.g. Walmsley et
al. 2000) suggesting that the W75~N nebula may exhibit similar
excitation conditions to those in Orion.

The \h\ and [FeII] line emission in W75~N does not conform
to what is expected for shock-excited emission resulting from the
interaction between a well-collimated jet and diffuse molecular gas.
In this respect, the physical characteristics of the W75~N flows
differ from their low-mass counterparts which produce collimated jets
observed in both \h\ and [FeII] emission.

Many previous authors have assumed that only VLA~1 (Ba) was in an
outflow stage based on the elongated morphology of the ionized gas,
the presence of \water\ maser emission along the UC HII region axis,
and because the position angles of the ionized gas and the CO emission
were similar.  \water\ maser emission is also associated with VLA~2
and VLA~3~(Bb) as well as MM~3 and MM~2, however, outflowing material
could not be traced to specific sources.  Assuming a single primary
driving source for the CO gas, Davis et al. (1998b) suggested that the
south-west CO red-shifted lobe and \h\ morphology supports a bow-shock
entrainment scenario for a molecular outflow driven by a jet from a
single massive star.  Our CO and millimeter continuum observations do
not support this theory that a single source drives the high-velocity
CO gas.  Further, our infrared observations suggest that the W75~N
outflows are not likely to be scaled-up versions of jet-driven outflows
from low-mass protostars.  

A question remains unanswered by this work: what source powers the
3~pc flow at P.A. 62.5\deg?  The flow mass is $~\gax~100$~M\sun, the
dynamical age is roughly $10^5$~years, and the mass loss rate $\dot M_f
\sim 10^{-4}$ to $10^{-3}$~M$_\odot$~yr$^{-1}$.  The flow parameters
are consistent with those produced by an early B protostar.
Hutawarakorn et al. (2002) have modeled the OH maser position-velocity
data and find evidence for a massive disk centered on VLA~2 (M$_{disk}
\sim 120$~M\sun\ with P.A. = 155\deg, roughly perpendicular to the
outflow axis).  A high-velocity, time-variable OH maser cluster is
coincident with VLA~2 suggesting an outflow origin.  Further, recent
observations with the Very Long Baseline Array (VLBA) show that a
clump of strong {\water} maser emission with high velocity dispersion
is centered on VLA~2 (Torrelles et al. 2003).  Thus, the OH and
{\water} maser activity suggests VLA~2 is producing a powerful
outflow.  Although our observations did not have adequate resolution
to isolate high-velocity gas toward VLA~2, we have determined that
VLA~1~(Ba), VLA~3~(Bb), and MM~2 are not likely to drive the 3~pc flow
that dominates the region dynamics.  Thus, it is possible that VLA~2
may drive the large-scale flow.  Follow-up observations at a
resolution less than $1''$ will be required to determine if, in fact,
VLA~2 drives the 3~pc flow.

Based on the size and velocity of the CO outflows from the W75~N (B) UC
HII regions, the region is $> 10^5$~years old.  W75~N (A) is more
evolved than the sources in MM~1 and the exciting star of W75~N (A)
has no detectable high-velocity gas associated with it.  The star,
detected in the infrared, is centered within a shell of warm dust
emission and an extended HII region (Hashick et al. 1981).  Figure 10
shows a color-color diagram using data from the 2MASS Point Source
Catalog for stars within $1'$ of MM~1 that were detected at all three
bands.  The locus of main-sequence stars is represented by the thick,
curved line (Bessell \& Brett 1988, Koornneef 1983) while the two
diagonal lines show reddening vectors up to $A_V = 40$ of dust
(adopting the $R_V = 5$ extinction law from Cardelli, Clayton, \&
Mathis 1989).  Sources within the reddening vectors have colors
consistent with main-sequence stars reddened by foreground dust.
Those to the right of the reddening vectors demonstrate excess
emission at $2\mu$m, consistent with the presence of circumstellar
material.  The infrared colors of the W75~N (A) exciting star are
consistent with those of a main-sequence star reddened by foreground
dust.  In comparison, the two bright stars to the south-east and
south-west of MM~1 (IRS~2 and IRS~3) have excess emission at $2\mu$m
consistent with the presence of circumstellar material.  The
protostars within MM~1 and MM~2 are not detectable at infrared
wavelengths.  W75~N represents a region of clustered star
formation which appears to be forming mid to early-B stars which exist
at a range of developmental stages.

VLA~1 (Ba) appears to have a well-collimated, outflow based on the
ionized gas morphology imaged by Torrelles et al. (1997) and the
presence of a relatively well-collimated, red-shifted CO lobe which
extends about 0.5~pc north-east of the source.  However, the spectral
type of the protostar is unknown since the ionized gas appears to be
due to thermal jet emission.  The well-collimated outflow which
appears to be produced by the embedded source in MM~2 is not detected
at centimeter wavelengths.  Either the powering source is not an
early-B star (e.g. it does not have sufficient ionizing radiation to
produce a detectable UC HII region) or accretion onto the protostar is
sufficiently high that it prevents the formation of a UC HII region
(see e.g. Churchwell 1999 and references therein).

There is no evidence for well-collimated flows (collimation ratios,
length/width, $>~10$) from the remaining embedded sources (early-B
protostars) or in the large-scale morphology of the CO, \h, \& [FeII]
emission.  The lack of highly-collimated flows from the known, early-B
protostars in W75~N suggests that it may be difficult for massive
stars to collimate outflowing material.  Although a few mid to early-B
protostars appear to be powering ionized jets, their molecular
outflows tend to be complex and poorly collimated (see, e.g., the
review by Shepherd 2002 and references therein).  To our knowledge,
there is no well-collimated {\it molecular} outflow powered by a
massive protostar (spectral type early-B to O) and most do not appear
to have ionized outflow components that are well-collimated
(e.g. Ridge \& Moore 2001, Shepherd, Claussen, \& Kurtz 2001;
Churchwell 1999).  Poorly collimated flows could be due to several
factors:

\vspace{-3mm}
\begin{itemize}
\item Confusion from multiple outflow sources in a cluster
      (e.g. W75~N: this work; or DR~21: Garden et al. 1991);

\vspace{-2mm}
\item Large flow precession angles (e.g. PV Ceph: Reipurth, Bally, 
      \& Devine 1997, Gomez, Kenyon, \& Whitney 1997; or IRAS
      20126$+$4104: Shepherd et al. 2000); 

\vspace{-2mm}
\item The presence of a strong wide-angle wind (e.g. Orion I:
      Greenhill et al. 1998; or G192.16--3.82: Shepherd, Claussen \&
      Kurtz 2001); and/or

\vspace{-2mm}
\item The molecular flow represents only the truncated base of
      a much larger flow (e.g. HH~80--81: Yamashita et al. 1989,
      Rodr\'{\i}guez et al. 1994; or G192.16--3.82: Devine et al. 1999).
\end{itemize} 

\vspace{-4mm}
\noindent
Our observations of W75~N supports the interpretation that massive
protostars may not be able to produce well-collimated molecular outflows.
This conclusion does not rule out the possibility that an underlying,
neutral jet may still exist as part of the outflow from the massive
protostars in W75~N.  High-resolution observations in shock tracers
such as SiO(J=1--0,v=0) or SiO(J=2--1,v=0) may be able to determine
whether collimated neutral jets are present in W75~N.

\section{SUMMARY}

W75~N represents an example of clustered, massive star
formation.  The cluster covers a wide range of evolutionary stages;
from stars with no apparent circumstellar material to deeply embedded
protostars actively powering massive outflows.  The CO outflow
measures more than 3~pc from end-to-end and is produced by at least
four individual sources.  \h\ emission extends well-beyond the CO
boundaries while [FeII] emission is only located close to the
protostellar cluster.  The CO, \h\ \& [FeII] morphology does not conform
to what is expected for shock-excited emission resulting from the
interaction between a well-collimated jet and diffuse molecular gas.
The irregular morphology of the infrared reflection nebula with
fingers of nebulosity radiating out from the millimeter cores supports
the conclusion that multiple energetic outflows are carving large
cavities in the molecular cloud.  More than 10\% of the molecular
cloud is outflowing material and the combined outflow energy is
roughly half the gravitational binding energy of the cloud.  Thus, the
observed W75~N outflows are injecting a significant amount of
mechanical energy into the cloud core and may help prevent further
collapse of the cloud.

\vspace{5mm}
\acknowledgements 
Research at the Owens Valley Radio Observatory is supported by the
National Science Foundation through NSF grant number AST 99-81546.
Star formation research at Owens Valley is also supported by NASA's
Origins of Solar Systems program, Grant NAGW-4030, and by the Norris
Planetary Origins Project.

D. P. Stark acknowledges support from the National Science Foundation
Research Experience for Undergraduate program.  

This paper is partly based on observations made with the Italian
Telescopio Nazionale Galileo (TNG) operated on the island of La Palma
by the Centro Galileo Galilei of the INAF (Istituto Nazionale di
Astrofisica) at the Spanish Observatorio del Roque de los Muchachos of
the Instituto de Astrofisica de Canarias.  The ARNICA and NICS
observations were performed in service mode by the TNG staff, we
especially acknowledge the help of Francesca Ghinassi, Juan Carlos
Guerra, and Antonio Magazz\'u.

This publication makes use of data products from the Two Micron All
Sky Survey, which is a joint project of the University of
Massachusetts and the Infrared Processing and Analysis
Center/California Institute of Technology, funded by the National
Aeronautics and Space Administration and the National Science
Foundation.


\clearpage

\begin{table}
\caption[]{W75~N 2.7~mm continuum emission}
\label{tab:par}
\smallskip
\begin{tabular}{|lcccc|}
\hline
 		&
		& Peak  	& Total 	&  \\
 		& Position
		& Flux Density  & Flux Density	& M$_{(gas+dust)}$ $^{\dagger}$ \\
Source 		& (J2000)
		& (mJy)		& (mJy)		& (M$_\odot$) \\
\hline
MM~1-4~~~~	& \ra{20}{38}{36.36}~~~\dec{$+$42}{37}{33.5}
		& 266   	& 650  		&  340 $\pm~ 70$ \\
MM~5	 	& \ra{20}{38}{37.78}~~~\dec{$+$42}{37}{59.0}
		& 51		& 129 		&  68  $\pm~ 16$ \\
MM~6	 	&\ra{20}{38}{36.31}~~~\dec{$+$42}{37}{55.9}
		& 25		& 38 		&  20  $\pm~ 6$ \\
MM~7	 	&\ra{20}{38}{36.56}~~~\dec{$+$42}{38}{12.7}
		& 25		& 43  		&  22  $\pm~ 6$ \\
MM~8	 	&\ra{20}{38}{33.68}~~~\dec{$+$42}{38}{01.7}
		& 23		& 31  		&  16  $\pm~ 5$ \\
MM~9	 	&\ra{20}{38}{38.70}~~~\dec{$+$42}{38}{19.8}
		& 19  		& 21  		&  11  $\pm~ 4$ \\
\hline
\end{tabular}

\vspace{.1in}
~~{\small $^{\dagger}$ Uncertainty includes $\pm 2$~M$_\odot$ due to image RMS
plus 20\% uncertainty in the absolute flux calibration.}
\end{table}

\begin{table}
\caption[]{W75~N Outflow Parameters}
\label{tab:par}
\smallskip
\begin{tabular}{|lll|}
\hline
Source:		    & MM~1 combined flows	
		    & MM~2 \\
\hline
\hline
CO radius of outflow  
                    &1.8 pc    &$>0.5$ pc         \\
Assumed inclination angle  
                    &45\deg\   &45\deg\              \\
Outflow Mass$^{\dagger}$ :       
                    &          &              \\
~~Western outflow                        
                    &68 M$_\odot$             
                    & \nodata\             \\
~~Eastern outflow 
                    &\underline{97 M$_\odot$} 
                    &\underline{$> 90$ M$_\odot$} \\
~~
                    & 165 M$_\odot$           
                    & $> 90$ M$_\odot$         \\
Momentum 
                    &$2.2 \times 10^3$ M$_\odot$ km s$^{-1}$  
                    &$> 1.1 \times 10^3$ M$_\odot$ km s$^{-1}$  \\
Kinetic Energy     
                    &$3.4 \times 10^{47}$ ergs    
                    &$> 1.8 \times 10^{47}$ ergs  \\
Dynamical time scale 
                    &$1.5 \times 10^5$ yr      
                    &$> 3.8 \times 10^4$ yr   \\
$\dot M_f$ 
                    &$1.2 \times 10^{-3}$ M$_\odot$ yr$^{-1}$ 
                    &$< 2.3 \times 10^{-3}$ M$_\odot$ yr$^{-1}$   \\
Momentum Supply Rate (Force) 
                    &$1.8 \times 10^{-2}$ M$_\odot$ km s$^{-1}$ yr$^{-1}$ 
                    &$< 2.9 \times 10^{-2}$ M$_\odot$ km s$^{-1}$ yr$^{-1}$ \\
Mechanical Luminosity 
                    & 23 L$_\odot$             
                    & $< 38$ L$_\odot$ \\
\hline
\end{tabular}

\vspace{3mm}
~~{\small $^{\dagger}$ MM~1 eastern outflow emission measured at
	velocities 2.2 to 4.8\kms\ and 15.2 to 36\kms. \\

\vspace{-5mm}
~~~~~ MM~1 western outflow emission measured at velocities --8.2 to
	4.8\kms\ and 15.2 to 36.0\kms.\\

\vspace{-5mm}
~~~ MM~2 outflow emission measured between --26.4 and 2.2\kms.}
\end{table}

\clearpage
\begin{center}
{\bf Figure Captions}
\end{center}
\noindent
{\bf Figure~1.} Integrated CO red-shifted (red lines) and blue-shifted
(blue lines) emission contours from 36.0\kms\ to 17.8\kms\ and
--0.4\kms\ to --26.4\kms, respectively. The images have an RMS of
23.5\jybkms\ with a peak of 56.0\jybkms\ in the red-shifted emission
image and 100.6\jybkms\ in the blue-shifted emission image.  Contours
begin at 10\% of the peak emission and continue at increments of
20\%. The synthesized beam is $6.46'' \times 6.28''$ at P.A.
$-54.7^\circ$.  \h\ line $+$ continuum emission is shown as grey scale
displayed as the square root of the intensity with a peak
of $7.45 \times 10^{-13}$~erg~cm$^{-2}$s$^{-1}$arcsec$^{-2}$.
W75~N~(A) is located at position $\alpha(J2000) =$
\ra{20}{38}{38}~ $\delta(J2000) =$ \dec{42}{37}{59} and W75~N~(B) at
position $\alpha(J2000) =$ \ra{20}{38}{37}~ $\delta(J2000) =$
\dec{42}{37}{32}.  UC HII regions embedded in the core of MM~1 are shown as
filled triangles while the millimeter cores MM~2--9 are shown as open
circles.  The large open circle (MM~5) is coincident with the infrared
emission associated with W75~N~A.  The solid black line delineates the
boundaries of the CO mosaic.

\noindent
{\bf Figure~2.} Continuum-subtracted \h\ line mosaic of W75~N.  The \h\
image is displayed on a linear scale from --0.6 to a peak intensity of
$2.6 \times 10^{-15}$ erg cm$^{-2}$ s$^{-1}$.  Individual knots of \h\
emission are labeled NE--A through I, C--A \& B, and SW--A
through K. Features observed by Davis et al. (1998a,b) are identified by
solid lines, previously undetected \h\ features which are more diffuse
are shown as dashed lines.  Large dashed boxes outline the fields
shown in Fig. 3.  

\noindent
{\bf Figure~3.} Continuum-subtracted \h\ emission shown in grey scale
with [FeII] emission shown as contours.  The [FeII] contours begin at
$5 \sigma$ and continue with a spacing of $8 \sigma$.  The left panel
shows the central and north-east outflow regions. The strong, diffuse
[FeII] emission is coincident with the W75~N~(A) and (B) reflection
nebulae.  The right panel shows the south-west outflow.  [FeII]
contours coincident with point sources are due to imperfect continuum
subtraction.

\noindent
{\bf Figures 4a \& 4b.}  CO channel images at 2.6\kms\ spectral
resolution between 32.1 and $-27.7$\kms.  The central velocity is
indicated in the upper right of each panel.  The LSR velocity is
10~\kms.  The RMS is 0.13\jyb\ and the peak emission is 16.9\jyb.  In
the first 12 and last 12 panels, contours are plotted from $\pm 4$, 8,
12, 16, 20~$\sigma$ and continue with a spacing of $10~ \sigma$.  In
the central 6 panels (8.7 to --4.3\kms), contours begin at $\pm 10,
30~ \sigma$ and continue with a spacing of $20 \sigma$.  Panels at
velocities 32.1 \& --27.7\kms\ show the synthesized beam in the
bottom right corner ($6.46'' \times 6.28''$ at P.A. $-54.7^\circ$) and
a scale size of 0.9 pc.  The plus symbol in each panel represents the
location of the peak emission in MM~1.  No other emission was detected
outside of this velocity range.

\noindent
{\bf Figure 5.}  The bottom left image shows continuum emission at
2.7~mm.  No other continuum sources were detected within the mosaic
field.  The image has an RMS of 3.6\mjyb.  Contours begin at $\pm 3,
4, 5, 7, 10, 20 \sigma$ and continue with a spacing of $10 \sigma$.
The greyscale is plotted on a linear scale from 7.2 to 265\mjyb.  The
synthesized beam in the lower right corner is $7.29'' \times 7.13''$
at P.A. $-62.9^\circ$.  UC HII regions in the center of MM~1 are
identified by filled triangles, MM~2--4 are shown as open circles.  A
scale size of 0.15 pc is represented by a bar in the lower left
corner.  The bottom right inset shows 3.3~mm continuum emission
obtained with $\sim 2''$ resolution (Figure 1 from Shepherd 2001).
Upper panels show the MM~5 millimeter source compared with narrow-band
\h$+$continuum emission and wide-band 2.12$\mu$m emission in W75~N~(A)
(from Fig.  1).

\noindent
{\bf Figure 6.}  The relationship between compact, high velocity CO
emission, infrared emission, millimeter continuum peaks, and UC HII
regions in W75~N~(B).  Red- and blue-shifted CO emission (upper
panels) is plotted from 20.4 to 36.0\kms\ and --5.6 to --23.8\kms,
respectively.  The RMS in both images is 0.5\jybkms; contours are
plotted from --3, 2, 3, 4~$\sigma$ and continue at spacings of
1~$\sigma$.   Millimeter core positions for MM~2--4
are shown as filled circles, UC HII regions embedded in MM~1 are
represented as filled triangles.  Proposed outflows are identified by
arrows from UC HII regions VLA~1 (Ba), VLA~3~(Bb), and MM~2.  The
synthesized beam ($2.04'' \times 1.77''$ at P.A. $-80.5^\circ$)
is shown in the top right image and a scale size of
0.1 pc is represented by a bar in the top left image.  The bottom two
images show the K$_s$ and \h\ emission relative to the proposed
outflows.  Both images are displayed as the square root of the
intensity.  

\noindent
{\bf Figure 7.}  
{\bf Top:} Integrated emission (zeroth moment) from 17.8\kms\ to
36.0\kms.  The RMS in the image is 1.2\jybkms, contours are plotted
from 4 to $20 \sigma$ with increments of $2 \sigma$ and then from 20
to $45 \sigma$ with increments of $5 \sigma$.  Greyscale is
plotted from $3 \sigma$ to a peak of 56.0\jybkms.  Millimeter cores
are identified as filled circles, UC HII regions by filled triangles.
The synthesized beam in the bottom left corner is $6.46'' \times
6.28''$ at P.A. $-54.7^\circ$.
{\bf Bottom} Position-Velocity plot along the length of the
high-velocity, red-shifted outflow.  Contours are plotted at 2, 5, 10,
15, 20, 30, 40, 50, 70, \& 90\% of the peak.  

\noindent
{\bf Figure 8.}  
{\bf Top:} Integrated emission (zeroth moment) from --5.6\kms\ to
--26.4\kms.  The RMS in the image is 0.96\jybkms, contours are plotted
from 4 to $20 \sigma$ with increments of $2 \sigma$, greyscale is
plotted from $3 \sigma$ to a peak of 20.0\jybkms.  Millimeter cores
are identified as filled circles, UC HII regions by filled triangles.
The synthesized beam in the bottom right corner is $6.46'' \times
6.28''$ at P.A. $-54.7^\circ$.
{\bf Bottom} Position-Velocity plot along the length of the
high-velocity, blue-shifted outflow.  Contours are plotted at 3, 5, 7, 9,
20, 40, 60, 80, \& 100\% of the peak.  

\noindent
{\bf Figure 9.}  CO(J=1--0) optical depth as a function of velocity
based on single-dish observations with the Kitt Peak 12~m telescope.
Optical depth is derived for three positions within the outflow: in
the north-east lobe (Top); centered on the MM~1 core (Center); and in
the south-west lobe (Bottom).  CO emission was measured in each
channel image of the interferometer mosaic and an optical depth
correction was made to the mass estimate based on the location and
velocity of the emission.

\noindent
{\bf Figure 10.}  {\bf Left:} Three color image using data from The Two
Micron All Sky Survey (2MASS).  The image is $1.5'$ on a
side.  The J-band image at 1.25$\mu$m is shown as blue, H-band at
1.65$\mu$m is green, \& K$_s$-band at 2.17$\mu$m is red. The $+$
symbol represents the location of the MM~1 peak.  {\bf Right:} A
color-color diagram using data from the 2MASS Point Source Catalog for
stars within $1'$ of MM~1 that were detected at all three bands.  
The locus of main-sequence stars is represented by the thick, curved
line while the two diagonal lines 
show reddening vectors up to $A_V = 40$ of dust.


\begin{thebibliography}{}

\vspace{-1mm}
\bibitem[]{}
Baart, E.E., Cohen, R.J., Davies, R.D., Norris, R.P., Rowland,
P.R. 1986, MNRAS, 219, 145

\vspace{-1mm}
\bibitem[2001]{Bea01}
Baffa, C., Comoretto, G., Gennari, S., Lisi, F., Oliva, E.,
Biliotti, V., Checcucci, A., Gavrioussev, V., Giani, E.,
Ghinassi, F., Hunt, L. K., Maiolino, R., Mannucci, F.,
Marcucci, G., Sozzi, M., Stefanini, P., \& Testi, L.
2001, A\&A, 378, 722
 
\vspace{-1mm}
\bibitem[]{}
Beck-Winchatz, B., B\"{o}hm, K.-H., \& Noriega-Crespo, A.  1996, AJ,
111, 346

\vspace{-1mm}
\bibitem[]{}
Bessell, M. S., \& Brett, J. M. 1988, PASP, 100, 1134

\vspace{-1mm}
\bibitem[]{}
Cabrit, S. \& Bertout, C. 1990, ApJ, 348, 530

\vspace{-1mm}
\bibitem{}
Cabrit, S., Ferreira, J., \& Raga, A. C. 1999, A\&A, 343, L61


\vspace{-1mm}
\bibitem[]{}
Cardelli, J. A., Clayton, G. C., \& Mathis, J. S. 1989, ApJ, 345, 245

\vspace{-1mm}
\bibitem[]{}
Churchwell, 1999, in ``The Origin of Stars and Planetary Systems'',
ed. Charles J. Lada and Nikolaos D. Kylafis (Kluwer Academic
Publishers), p. 515


\vspace{-1mm}
\bibitem[]{}
Cornwell, T. \& Braun, R. 1988, in ``Synthesis Imaging in Radio
Astronomy,'' eds. R. A. Perley, F. R. Schwab \& A. H. Bridle, PASP, 
p 167

\vspace{-1mm}
\bibitem[]{}
Davis, C. J., Moriarty-Schieven, G. H., Eisl\"{o}ffel, J., Hoare, M. G.,
\& Ray, T. P. 1998a, AJ, 115, 1118

\vspace{-1mm}
\bibitem[]{}
Davis, C.J., Smith, M.D., \& Moriarty-Schieven, G.H. 1998b, MNRAS, 299,
825

\vspace{-1mm}
\bibitem[]{}
Devine, D., Bally, J., Reipurth, B., Shepherd, D., Watson, A. 1999,
AJ, 117, 2919

\vspace{-1mm}
\bibitem[]{}
Dickel, H. R., Wendker, H. J., \& Bieritz, J. H. 1969, A\&A, 1, 270

\vspace{-1mm}
\bibitem[]{}
Fischer, J., Sanders, D. B., Simon, M., \& Solomon, P. M. 1985, ApJ,
293, 508

\vspace{-1mm}
\bibitem[]{}
Garden, R. P., Hayashi, M., Gatley, I., Hasegawa, T., \& Kaifu,
N. 1991, ApJ, 374, 540

\vspace{-1mm}
\bibitem[2002]{Gea02}
Ghinassi, F., Licandro, J., Oliva, E., Baffa, C., Checcucci, A., 
Comoretto, G., Gennari, S., \& Marcucci, G. 2002, A\&A, 386, 1157

\vspace{-1mm}
\bibitem[]{}
Gomez, M., Kenyon, S. J., \& Whitney, B. A. 1997, AJ, 114, 265

\vspace{-1mm}
\bibitem[]{}
Gredel, R. 1994, A\&A, 292, 580

\vspace{-1mm}
\bibitem[]{}
Greenhill, L. J., Gwinn, C. R., Schwartz, C., Moran, J. M., \&
Diamond, P. J. 1998, Nature, 396, 650


\vspace{-1mm}
\bibitem[]{}
Haschick, A. D., Reid, M. J., Burke, B. F., Moran, J. M., \& Miller,
G. 1981, ApJ, 244, 76

\vspace{-1mm}
\bibitem[]{}
Hildebrand, R. H. 1983, Q.Jl.R.Astr.Soc, 24, 267

\vspace{-1mm}
\bibitem[]{}
Hollenbach, D. \& Mckee, C. F. 1989, ApJ, 342, 306

\vspace{-1mm}
\bibitem[1996]{Hea96}
Hunt, L.K., Lisi, F., Testi, L., Baffa, C., Borelli, S., Maiolino, R., 
Moriondo, G., \& Stanga, R. M. 1996, A\&AS, 115, 181

\vspace{-1mm}
\bibitem[1998]{Hea98}
Hunt, L.K., Mannucci, F., Testi, L., Migliorini, S., Stanga, R.M.,
Baffa, C., Lisi, F., \& Vanzi, L. 1998, AJ, 115, 2594

\vspace{-1mm}
\bibitem[]{}
Hunter, T.R., Taylor, G.B., Felli, M., \& Tofani, G. 1994, A\&A, 284,
215
 
\vspace{-1mm}
\bibitem[]{}
Hutawarakorn, B., Cohen, R. J., \& Brebner, G. C. 2002, MNRAS, 330, 349

\vspace{-1mm}
\bibitem{}
Kalenski\u{\i}, S. V., Promislov, V. G., Alakoz, A. V., Winnberg, A.,
\& Johansson, L. E. B. 2000, Astronomy Reports, 44, 725

\vspace{-1mm}
\bibitem{}
K\"{o}nigl, A. 1999, New Astronomy Reviews, 43, 67

\vspace{-1mm}
\bibitem{}
K\"{o}nigl, A. \& Pudritz, R. E. 2000, in ``Protostars and Planets
IV'', ed. V. Mannings, A. Boss \& S. Russell (Tucson: Univ. of Arizona
Press), in press 
 
\vspace{-1mm}
\bibitem[]{}
Koornneef, J. 1983, A\&A, 128, 84

\vspace{-1mm}
\bibitem[]{}
Kramer, C., Alves, J., Lada, C., Lada, E., Sievers, A., Ungerechts,
H., \& Walmsley, M. 1998, A\&A, 329, L33

\vspace{-1mm}
\bibitem[]{}
Lada, C. J. \& Fich, M. 1996, ApJ, 459, 638

\bibitem[1996]{Lea96}
Lisi, F., Hunt, L.K., Baffa, C., Biliotti, V., Bonaccini, D., 
del Vecchio, C., Gennari, S., Hunt, L. K.,
 Marcucci, G., \& Stanga, R. 1996, PASP, 108, 364

\vspace{-1mm}
\bibitem[]{}
Lorenzetti, D., Giannini, T., Vitali, F., Massi, F., \& Nisini,
B. 2002, ApJ, 564, 839

\vspace{-1mm}
\bibitem[]{}
Minier, V., Conway, J. E., \& Booth, R. S. 2000, A\&A, 362, 1093

\vspace{-1mm}
\bibitem[]{}
Minier, V., Conway, J. E., \& Booth, R. S. 2001, A\&A, 369, 278

\vspace{-1mm}
\bibitem[]{}
Moore, T. J. T., Mountain, C. M., Yamashita, T., \& McLean, I. S. 1991,
MNRAS, 248, 377

\vspace{-1mm}
\bibitem[]{}
Moore, T. J. T., Mountain, C. M., \& Yamashita, T. 1991, MNRAS, 248,
79
 
\vspace{-1mm}
\bibitem[]{}
Moore, T. J. T., Mountain, C. M., Yamashita, T., \& Selby, M. J. 1988,
MNRAS, 234, 95

\vspace{-1mm}
\bibitem[]{}
Nisini, B, Caratti o Garatti, A., Giannini, T. \& Lonenzetti, D. 2002,
A\&A, Accepted

\vspace{-1mm}
\bibitem[]{}
Pollack, J. B., Hollenbach, D., Beckwith, S., Simonelli, D. P., Roush,
T., \& Fong, W. 1994, ApJ, 421, 615

\vspace{-1mm}
\bibitem[]{}
Reipurth, B., Bally, J., \& Devine, D. 1997, AJ, 114, 2708

\vspace{-1mm}
\bibitem[]{}
Reipurth B., Yu, K. C., Heathcote, S., Bally, J., \& Rodr\'{\i}guez,
L. F. 2000, AJ, 120, 1449
  
\vspace{-1mm}
\bibitem[]{}
Richer, J. S., Shepherd, D. S., Cabrit, S., Bachiller, R., \&
Churchwell, E. 2000, in ``Protostars and Planets IV'', ed. V. Mannings,
A. Boss \& S. Russell (Tucson: Univ. of Arizona Press), 867

\vspace{-1mm}
\bibitem[]{}
Ridge, N. A. \& Moore, T. J. T. 2001, A\&A, 378, 495

\vspace{-1mm}
\bibitem[]{}
Rodr\'{\i}guez, L. F., Garay, G., Curiel, S., Ramirez, S., Torrelles,
J. M., Gomez, Y., \& Velazquez, A. 1994, ApJ, 430, L65

\vspace{-1mm}
\bibitem[]{}
Sault, R. J., Teuben, P. J., \& Wright, M. C. H. 1995, in
Astronomical Data Analysis Software and Systems IV, ed. R. A. Shaw,
H. E. Payne, \& J. J. E. Hayes, PASP Conf Series 77, 433

\vspace{-1mm}
\bibitem[]{}
Scoville, N. Z., Sargent, A. I., Sanders, D. B., Claussen, M. J.,
Masson, C. R., Lo, K. Y., \& Phillips, T. G. 1986, ApJ, 303, 416
 
\vspace{-1mm}
\bibitem[]{}
Scoville, N. Z., Carlstrom, J. E., Chandler, C. J., Phillips, J. A.,
Scott, S. L., Tilanus, R. P. J., \& Wang, Z. 1993, PASP, 105, 1482
 
\vspace{-1mm}
\bibitem{}
Shang, H., Glassgold, A. E., Shu, F. H., Lizano, S. 2002, ApJ, 564,
853


\vspace{-1mm}
\bibitem[]{}
Shepherd, D. S., Yu, K. C., Bally, J., \& Testi, L. 2000, ApJ, 535,
833 

\vspace{-1mm}
\bibitem[]{}
Shepherd, D. S., Claussen, M. J., \& Kurtz, S. E. 2001, Science, 292,
1513 

\vspace{-1mm}
\bibitem[]{}
Shepherd, D. S. 2001, ApJ, 546, 345

\vspace{-1mm}
\bibitem[]{}
Shepherd, D. S. \& Watson, A. M. 2002, ApJ, 566

\vspace{-1mm}
\bibitem[]{}
Shepherd, D. S. 2002, in ``Galactic Star Formation Across the Stellar
Mass Spectrum,'' ASP conference series, in press

\vspace{-1mm}
\bibitem{}
Shu, F. H., Najita, J. R., Shang, H., \& Li, Z.-Y.  2000, in
Protostars and Planets IV, ed. V. Mannings, A. Boss \& S. Russell
(Tucson: Univ. of Arizona Press), 789

\vspace{-1mm}
\bibitem{}
Slysh, V. I., Migenes, V., Val'tts, I. E., Lyubchenko, S. Yu.,
Horiuchi, S., Altunin, V. I., Fomalont, E. B., \& Inoue, M. 2002, ApJ,
564, 317

\vspace{-1mm}
\bibitem[]{}
Smith, M. D. 1994, A\&A, 289, 256

\vspace{-1mm}
\bibitem[]{}
Smith, M. D., Suttner, G., \& Yorke, H. W. 1997, A\&A, 323

\vspace{-1mm}
\bibitem[]{}
Suttner, G., Smith, M. D., Yorke, H. W., \& Zinnecker, H. 1997, A\&A,
318 

\vspace{-1mm}
\bibitem[]{}
Torrelles, J.M., Gomez, J.F., Rodr\'{\i}guez, L.F., Ho, P.T.P.,
Curiel, S., \& Vazquez, R. 1997, ApJ, 489, 744

\vspace{-1mm}
\bibitem[]{}
Torrelles, J.M., et al. 2003, in preparation

\vspace{-1mm}
\bibitem[1998]{Vea98}
Vanzi, L., Gennari, S., Ciofini, M., \& Testi, L. 1998,
Exp. Astronomy, 8, 177 

\vspace{-1mm}
\bibitem[]{}
Walmsley, C. M., Natta, A., Oliva, E., \& Testi, L. 2000, A\&A, 364,
301 

\vspace{-1mm}
\bibitem[]{}
Watson, C., Churchwell, E., \& Pankoni, V., 2002, ApJ, in press

\vspace{-1mm}
\bibitem[]{}
Wilson, T. L., \& Rood, R. T. 1994, ARAA, 32, 191

\vspace{-1mm}
\bibitem[]{}
Yamashita, T., Suzuki, H., Kaifu, N., Tamura, M., Mountain, C. M.,
Moore, T. J. T. 1989, ApJ, 347, 894
\end{thebibliography}
\end{document}